\documentclass[a4paper]{jpconf_mod}
\usepackage{graphicx}
\usepackage{amsmath,xfrac,paralist}
\usepackage{url}

\def\tev{\textrm{TeV}}
\def\gev{\textrm{GeV}}

\def\ipb{\ensuremath{\textrm{pb}^{-1}}}
\def\ifb{\ensuremath{\textrm{fb}^{-1}}}

\def\to{\ensuremath{\rightarrow}}

\def\X{\ensuremath{\tilde\chi_1^0}}
\def\R{\emph{R}}

\def\pt{\ensuremath{p_{\rm T}}}
\def\met{\ensuremath{E_{\rm T}^{\rm miss}}}
\def\mef{\ensuremath{M_{\rm eff}}}

\def\mt{\ensuremath{M_{\rm T}}}

\begin{document}
\title{Experimental status of particle and astroparticle searches for supersymmetry}

\author{Vasiliki A Mitsou}

\address{Instituto de F\'isica Corpuscular (IFIC), CSIC -- Universitat de Val\`encia, \\
Parc Cient\'ific de la U.V., C/ Catedr\'atico Jos\'e Beltr\'an 2,
E-46980 Paterna (Valencia), Spain}

\ead{vasiliki.mitsou@ific.uv.es}

\begin{abstract}
An overview of supersymmetry searches is presented, covering collider experiments, direct and indirect searches for supersymmetric dark matter. Recent LHC experimental results are reviewed, and the constraints from $B$-meson decays are reported. Implications for supersymmetry of the latest direct and indirect searches are thoroughly discussed. The focus is on the complementarity of the various probes --- particle and astrophysical --- for constraining Supersymmetry.
\end{abstract}

\section{Introduction}\label{sc:intro}

The Standard Model provides the current most accurate description of elementary particle physics. It has been experimentally tested up to the TeV scale with remarkably successful results. Nevertheless, there are pieces of evidence pointing to physics beyond the SM such as the existence of dark matter (DM), the matter-antimatter asymmetry, the neutrino masses, and the hierarchy problem. Several theories have been proposed to address these issues and Supersymmetry (SUSY)~\cite{susy-intro} is one of the most favoured. 

SUSY postulates a discrete symmetry between fermions and bosons, thus assigning a new particle, a \textit{superpartner}, to each SM field. Since no SUSY particles have been found yet, SUSY must be a broken symmetry, leading to numerous proposed models and various symmetry-breaking mechanisms. SUSY is characterised by a large number of free parameters, e.g.\ more than 100~parameters are introduced in the   Minimal Supersymmetric Standard Model (MSSM). Nonetheless, a variety of experimental and observational handles are at our disposal to search for SUSY and constrain it. The different facets of SUSY manifestation and the current status of the hunt for SUSY is discussed hereupon.

This paper is structured as follows. Section~\ref{sc:dm} provides a brief introduction to dark matter and its implications for SUSY, also discussing alternative SUSY models yielding a non-trivial DM density. It reviews both direct and indirect DM detection methods and recent results. In Section~\ref{sc:searches} the features of collider experiments that play a central role in exploring SUSY are highlighted. There we concisely discuss the SUSY constraints originating from $B$-physics experiments, whilst strong emphasis is given on recent attempts to discovering supersymmetry at the Large Hadron Collider (LHC)~\cite{lhc} and the ensuing exclusion limits, covering both \R-parity conserving and \R-parity breaking scenarios --- the latter receiving special attention. The implications for SUSY arising from the observation of a Higgs-like particle are also addressed. The paper concludes with a summary in section~\ref{sc:summary}.

\section{SUSY as dark matter}\label{sc:dm}

\begin{figure}[ht]
\begin{minipage}[b]{0.48\textwidth}
\includegraphics[width=\textwidth]{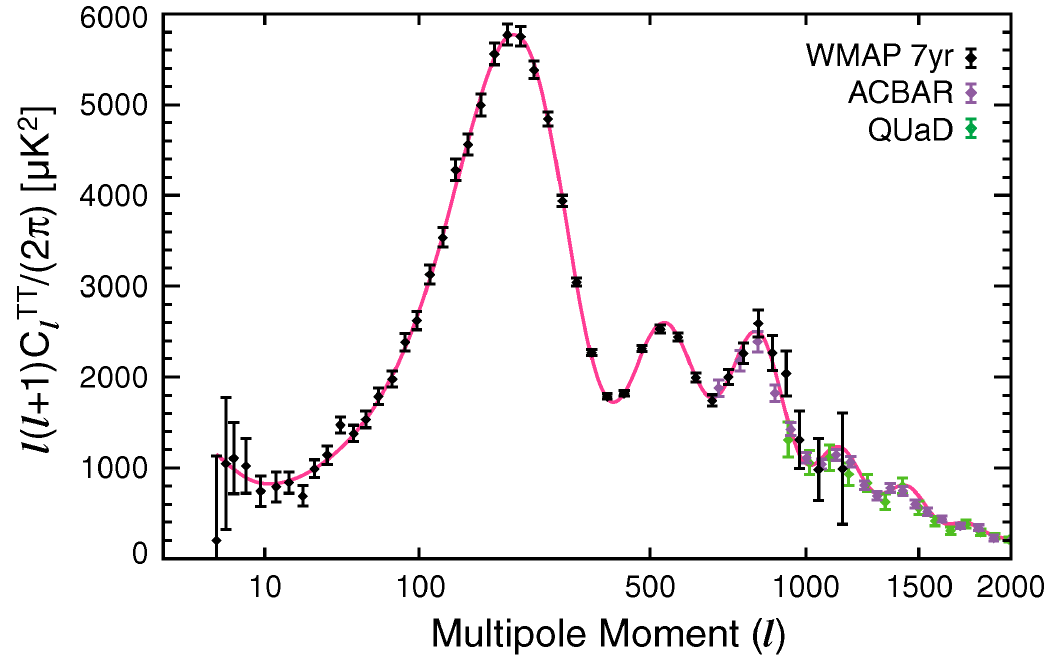}
\caption{\label{fg:wmap}The WMAP 7-year temperature power spectrum, along with the temperature power spectra from the ACBAR and QUaD experiments. The solid line shows the best-fitting 6-parameter flat CDM model to the WMAP data alone. From~\cite{wmap}.}
\end{minipage}\hspace{0.04\textwidth}%
\begin{minipage}[b]{0.48\textwidth}
\includegraphics[width=\textwidth]{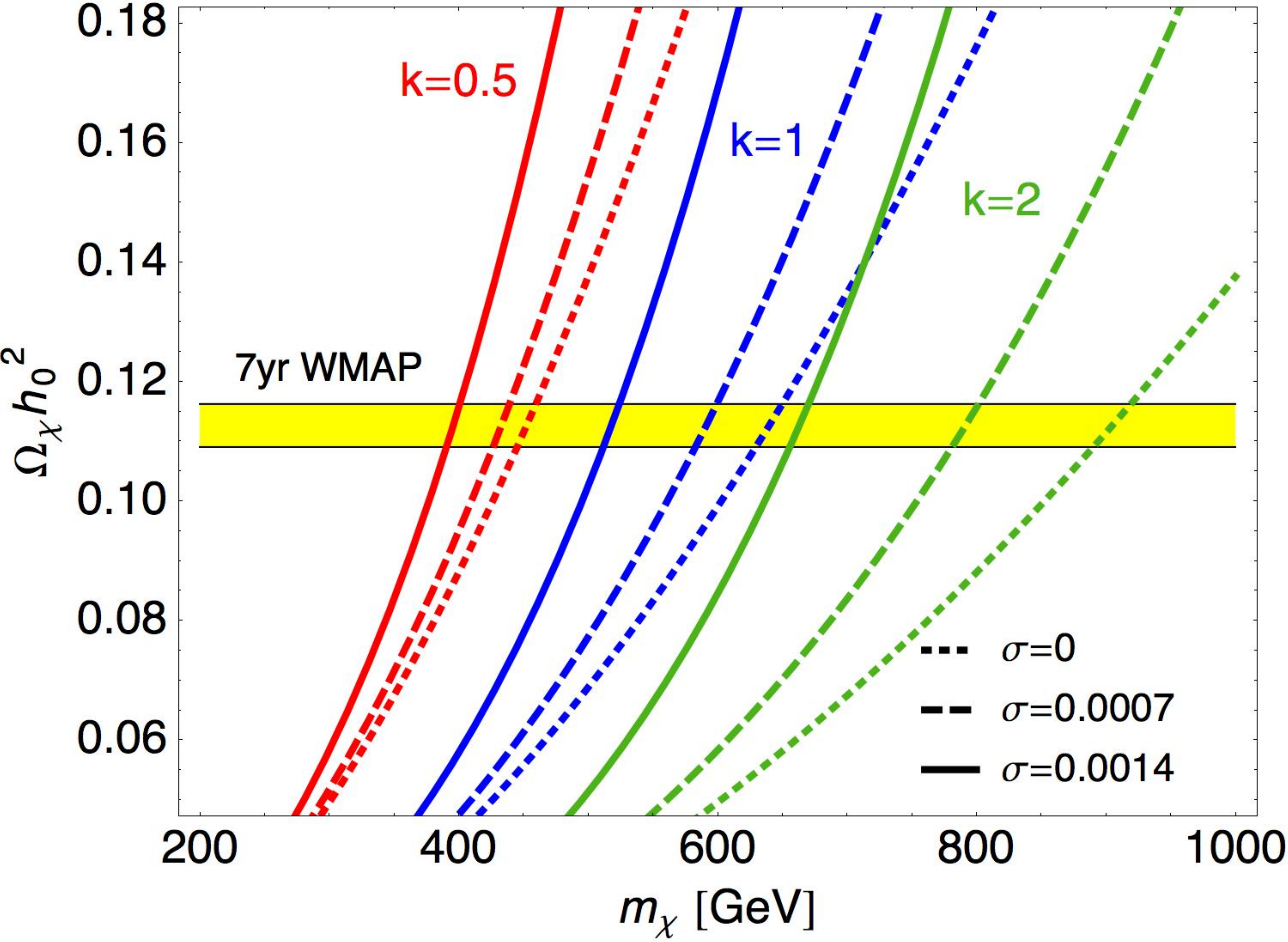}
\caption{\label{fg:omegaxh2}Predicted values of $\Omega_{\chi}h^2$ as a function of the DM particle mass $m_{\chi}$ for various deviations $k$ from the weak coupling and for different values of foam fluctuations $\sigma$. The yellow horizontal band represents the WMAP7 value for $\Omega_{\chi}h^2$~\cite{wmap}. From~\cite{finsler}.}
\end{minipage} 
\end{figure}

Unveiling the nature of dark matter~\cite{dm-review,vaso-dm,cirelli} is a quest in both Astrophysics and Particle Physics.  According to observations over the past two decades --- obtained by combining a variety of astrophysical data, such as type-Ia supernovae~\cite{snIa}, cosmic microwave background (CMB)~\cite{wmap}, baryon oscillations~\cite{bao} and weak lensing data~\cite{lensing} ---, most of our Universe energy budget consists of unknown entities: $\sim\!23\%$ is dark matter and $\sim\!72\%$ is dark energy, a form of ground-state energy. Dark matter existence is inferred from gravitational effects on visible matter, but is undetectable by emitted or scattered electromagnetic radiation. The most precise measurement comes from anisotropies of the cosmic microwave background~\cite{wmap,cmb}; the third peak in the temperature power spectrum, shown in Fig.~\ref{fg:wmap}, is used to extract information about dark matter.

Weakly-interacting massive particles (WIMPs), where the typical (weak-scale) annihilation cross sections are of the same order of magnitude as the thermally-averaged DM annihilation cross section, constitute the most popular class of dark matter candidates. Supersymmetry is a theoretical scenario that inherently proposes such dark matter candidates. Within SUSY with \R-parity conservation, the particle that plays the role of dark matter is the lightest supersymmetric particle (LSP); LSPs that have the right properties  are the axino, the gravitino (as superWIMPs) and the lightest neutralino (as WIMP). Such scenarios --- should they indeed explain dark matter --- are severely constrained from the Big-Bang nucleosynthesis, the measured DM abundance and the results of direct DM detection experiments and indirect detection telescopes.  

\subsection{DM relic density and SUSY}\label{sc:relic}

WIMPs were created thermally in the early universe, hence their relic abundance, $\Omega_{\chi}$, and their properties, such as mass and couplings, are closely related through the averaged annihilation cross section. For a specific SUSY model, the predicted neutralino relic density, calculated by taking into account all possible LSP annihilation processes, is compared to the measured $\Omega_{\chi}$, imposing constraints on the model parameters. 

The DM relic density estimation is based on the Standard Cosmological Model ($\Lambda$CDM)~\cite{lcdm}, involving cold DM as the dominant DM species, and a positive cosmological constant $\Lambda>0$. Nevertheless, the possibility for different theoretical scenarios is open, that modify the estimated DM relic abundance for given cosmological observations. For instance, the presence of space-time ($D$-particle) foam in string/brane-theory leads to Finsler-type metric distortions, induced by interactions of the DM particle(s) with the defects in the foam~\cite{finsler}. These metrics lead to modifications in the pertinent Boltzmann equation and consequently to enhancement of the relic abundances, as displayed in Fig.~\ref{fg:omegaxh2}. However stringent constraints on the defect density in $D$-foam models can also be imposed by astrophysical telescopes on the arrival times of high energy cosmic photons~\cite{grb}.  

The presence of the time-dependent dilaton may also affect the dark matter relic density calculation, since it modifies the Boltzmann equations. In this case, a \emph{dilution} in the neutralino density of ${\cal O}(10)$ is predicted~\cite{dutta}, widening the allowed parameter space of supersymmetry at collider searches. Furthermore, other astrophysical observations such as type-Ia supernovae and galaxy ages set tight constraints on the parameters of these models through their prediction for the dark energy contribution~\cite{nonequil-observ}.

\subsection{Direct WIMP detection}\label{sc:direct}

In direct detection low background underground experiments, one attempts to observe the nuclear recoil produced by WIMP scattering off nucleons~\cite{direct}. The expected signal features a recoil spectrum which falls exponentially with energy and extends to a few tens of keV only. Various techniques applied to observe the recoil with uncorrelated systematic uncertainties: scintillator NaI, cryogenic, noble liquids (Ar, Xe), bubble chamber (superheated liquid).

\begin{figure}[ht]
\centering
\includegraphics[width=0.65\textwidth]{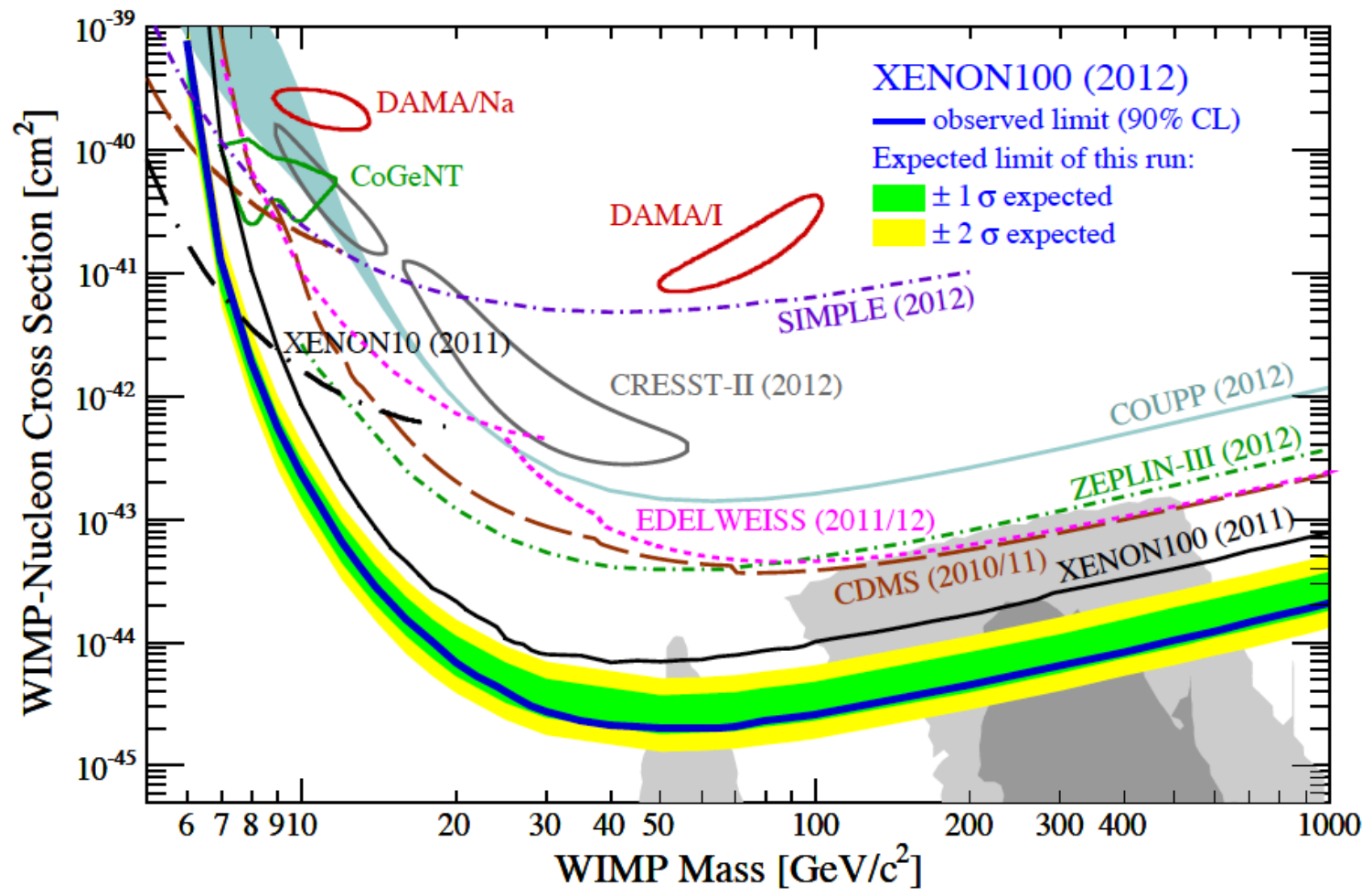}%
\caption{\label{fg:xenon} Spin-independent WIMP-nucleon scattering from XENON100: The expected sensitivity is shown by the green/yellow band $(1\sigma/2\sigma)$ and the resulting exclusion limit (90\% CL) in blue. From~\cite{xenon100}.}
\end{figure}

The current experimental landscape is shown in Fig.~\ref{fg:xenon} in terms of WIMP-nucleus elastic cross section versus WIMP mass~\cite{xenon100}. The positive results shown represent the annual modulation observed by DAMA/LIBRA~\cite{dama} and CoGeNT~\cite{cogent}, as well as unidentified excesses of events seen by CoGeNT~\cite{cogent} and CRESST-II~\cite{cresst}. CDMS does not see any hint of annual modulation~\cite{cdms} although their energy threshold is higher than CoGeNT. The XENON100 data~\cite{xenon100} lead to the strongest exclusion limit so far for constraining the DM-nucleon cross section to below $10^{-44}~{\rm cm}^2$ for $m_{\chi}\sim100~\gev$. For comparison, other experimental results are also shown~\cite{xenon-other,cogent,cresst}, together with the $1\sigma/2\sigma$ regions preferred by supersymmetric (Constrained MSSM) models~\cite{xenon-cmssm}, still untouched by the XENON limits. 

It is worth noting that these results strongly depend on astrophysical uncertainties, such as the local DM density and DM velocity distribution and on assumptions made regarding the WIMP interaction with the detector active material. Finally, systematic uncertainties in the nuclear modeling can affect the comparison between different detection techniques. 

\subsection{Indirect searches for DM}\label{sc:indirect}

WIMPs can annihilate and their annihilation products can be detected; these include neutrinos, $\gamma$-rays, positrons, antiprotons, and antinuclei~\cite{indirect}. ``Smoking gun'' signals for indirect detection are neutrinos coming from the center of the Sun or Earth, and mono-energetic photons from WIMP annihilation in space. These techniques could distinguish between different coupling scenarios and the nature of WIMPs: neutralinos, Kaluza-Klein states, etc.              
 
\begin{figure}[ht]
\centering
\includegraphics[width=0.7\textwidth]{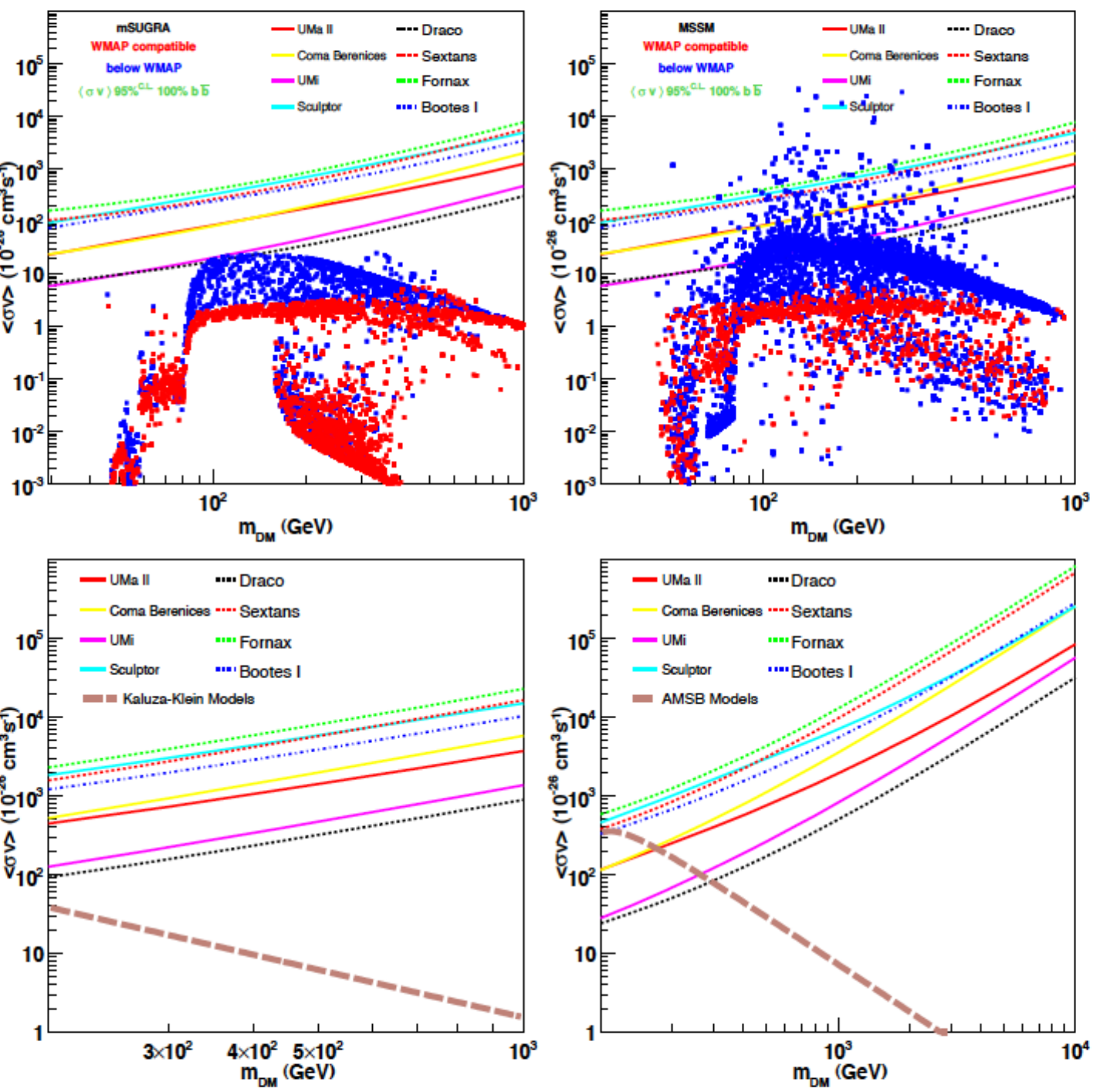}%
\caption{\label{fg:fermi} mSUGRA (upper left), MSSM (upper right), Kaluza-Klein UED (lower left) and anomaly mediated (lower right) models in the $(m_\text{WIMP},\langle\sigma v\rangle)$ plane. Red points have a $\Omega_{\chi}$ compatible with the inferred DM density (blue points have a lower DM density). The lines indicate the Fermi 95\% upper limits from likelihood analysis on the
selected dwarfs. From~\cite{fermi}.}
\end{figure}

The observation of 14~dwarf spheroidal galaxies with the Fermi Gamma-Ray Space Telescope showed no significant $\gamma$-ray emission above 100~MeV from the candidate dwarf galaxies~\cite{fermi}. Therefore upper limits to the $\gamma$-ray flux assuming representative spectra from WIMP annihilation were set, as demonstrated in Fig.~\ref{fg:fermi}, also compared with predictions from mSUGRA, MSSM with a reduced set of parameters~\cite{gondolo}, Kaluza-Klein dark matter in universal extra dimensions (UED) and wino-like dark matter in AMSB. The upper limits on the $\gamma$-ray flux are already competitive for MSSM models, provided that they correspond to low thermal relic density. Furthermore, these flux upper limits already disfavour AMSB models with masses $<300~\gev$. 

Controversial hints of WIMP annihilation have been seen so far in data by PAMELA~\cite{pamela}, EGRET, Fermi-LAT~\cite{fermi-excess}, which may be attributed to the large astrophysical uncertainties that enter in the data analysis. At present, much activity and excitement surrounds a tentative line signal at $E_{\gamma}\simeq135~\gev$~\cite{fermi-line}. The required annihilation cross section to explain this signal is very large, however supersymmetric models with such large signals do exist. Further progress to determine if the line signal is robust and to improve sensitivities for both continuum and line searches is sure to come from continued running of existing and upcoming experiments.  

\begin{figure}[ht]
\centering
\includegraphics[width=0.6\textwidth]{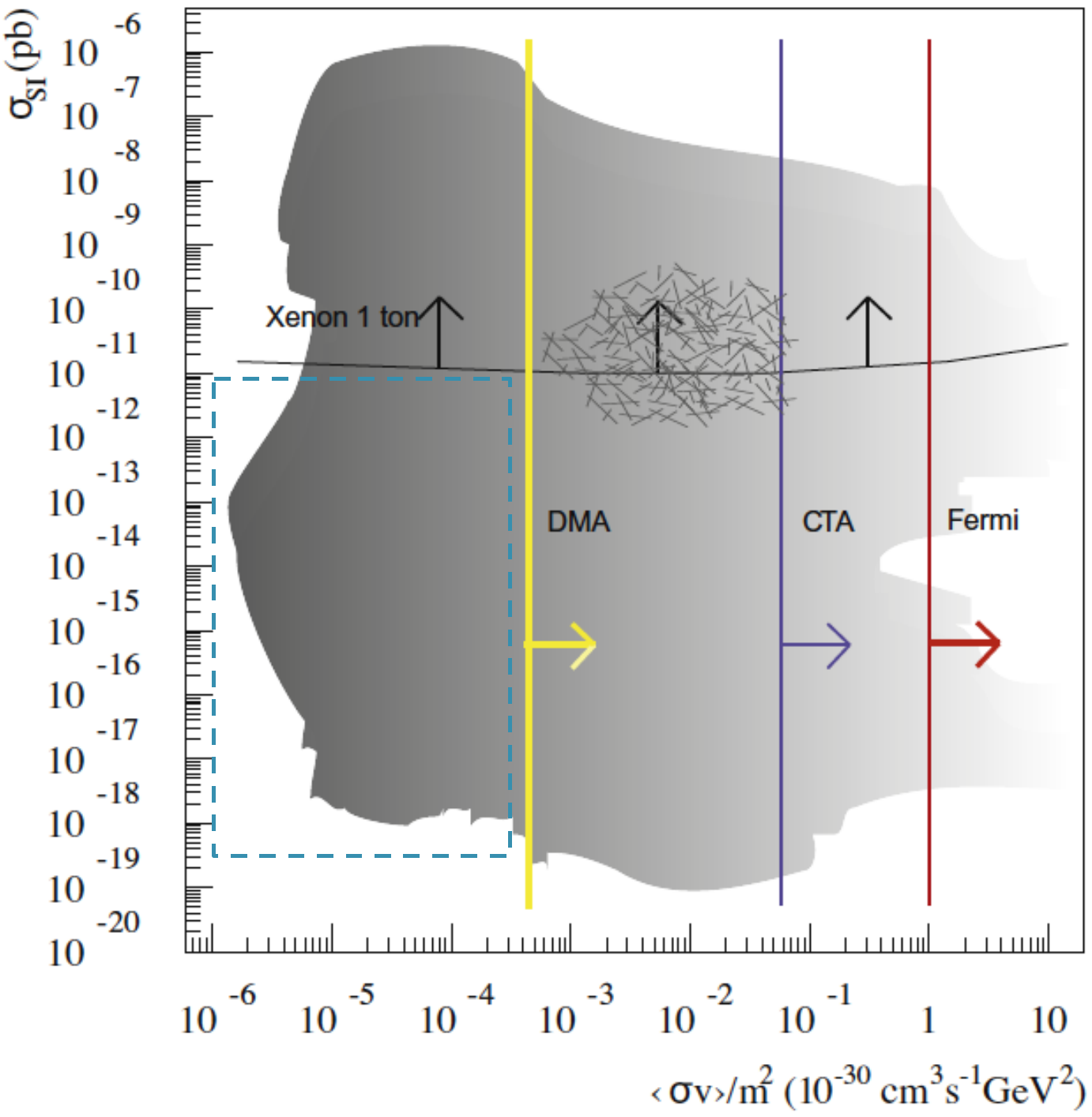}%
\caption{\label{fg:complementary} Illustration of the reach of direct and indirect dark matter detection experiments for $\gamma$-ray detection. The shaded region is the approximate range of WMAP-compatible MSSM model space. The smaller hatched region is a rough estimate of where Kaluza-Klein models reside~\cite{ued}. The dashed rectangular indicates the region that may be covered by LHC and neutrino telescopes. The MSSM results are based on Ref.~\cite{complementary}, to which the reader is directed for information on the mass range, composition, etc. From~\cite{bergstrom}.}
\end{figure}

To recapitulate, direct and indirect DM detection techniques set severe constraints on supersymmetric models, yet accelerator experiments are required to provide a measurement of different model parameters exploiting uncorrelated systematics. This is concisely depicted in Fig.~\ref{fg:complementary}, where the coverage of the direct and indirect detection with photons are complemented by the LHC and neutrino telescopes~\cite{bergstrom}. In the following years we expect a continuous interplay between particle physics experiments and astrophysical/cosmological observations.

\section{Searches in colliders}\label{sc:searches}

The stringent exclusion limits set on supersymmetric models and parameters come from collider experiments, such as the LEP, Tevatron and LHC detectors. The LHC currently in operation at CERN in Geneva, Switzerland, is an ideal machine for discovering supersymmetry. Although here we focus on direct searches for SUSY signals, discussed in Section~\ref{sc:LHC}, SUSY is also constrained indirectly through the (non-)observation of rare decays of $B$~mesons.Ê 

\subsection{$B$-physics constraints}\label{sc:bphysics}

Precise experimental measurements and theoretical predictions have been achieved for the $B$-meson systems in the past decade and stringent constraints due to considerable beyond-SM contributions to many observables have been acquired~\cite{harnew}. The rare decay $BR(B_s \to \mu^+\mu^-)$ deserves special attention as new results have been recently announced by the LHCb collaboration~\cite{lhcb-det} using $pp$ collisions at $\sqrt{s}=8~\tev$. An excess of $B_s \to \mu^+\mu^-$ candidates with respect to the background expectation is observed with a $3.5\sigma$ significance~\cite{lhcb}. This result is in agreement with the Standard Model expectation, therefore limits on new-physics models are set.

Here, we review constraints obtained with the LHCb results using 1~\ifb of data at $\sqrt{s}=7~\tev$, where a stringent 95\%~CL limit on the branching ratio $BR(B_s \to \mu^+\mu^-) < 4.5\cdot10^{-9}$ has been obtained~\cite{lhcb-old}. The decay $B\to K^+\mu^+\mu^-$, on the other hand, provides a variety of complementary observables as it gives access to angular distributions in addition to the differential branching fraction. In addition to the above observables, $B \to X_s\gamma$, $B \to \tau\nu$, $B \to D\tau\nu_{\tau}$, $B \to X_s\mu^+\mu^-$ and $D_s \to\tau\nu_{\tau}$ are also very sensitive to SUSY.

A comparison between different flavour observables in the plane $(m_{1/2},\,m_0)$ is given in Fig.~\ref{fg:b-phys}, where limits from $B \to X_s\gamma$, $B \to \tau\nu$, $R_{l23}(K\to\mu\nu_{\mu})$, $B \to D\tau\nu_{\tau}$, $B \to X_s\mu^+\mu^-$ and $D_s \to \tau\nu_{\tau}$ are also shown~\cite{b-phys}. The LHCb limit strongly constrains the CMSSM with large $\tan\beta=50$, however for $\tan\beta=30$, as can be seen from the figure, the flavour constraints and in particular $B_s \to \mu^+\mu^-$, are inferior to those from direct searches. On the other hand, $B \to X_s\gamma$ and $B\to K^+\mu^+\mu^-$ related observables and in particular the forward-backward asymmetry could play a complementary role in the intermediate $\tan\beta$ regime.

\begin{figure}[ht]
\centering
\includegraphics[width=0.55\textwidth]{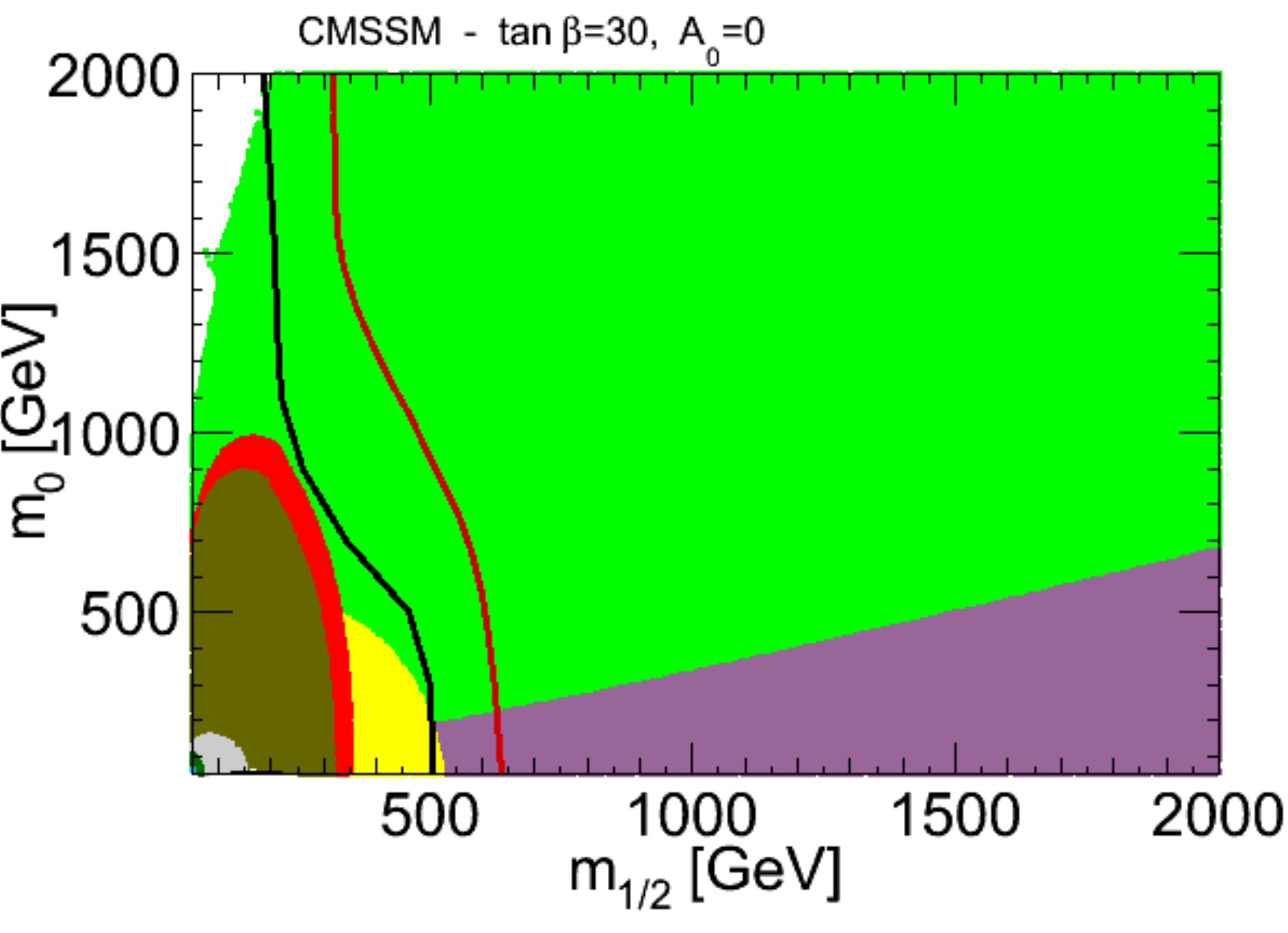}%
\hspace{2pc}%
\includegraphics[width=0.19\textwidth]{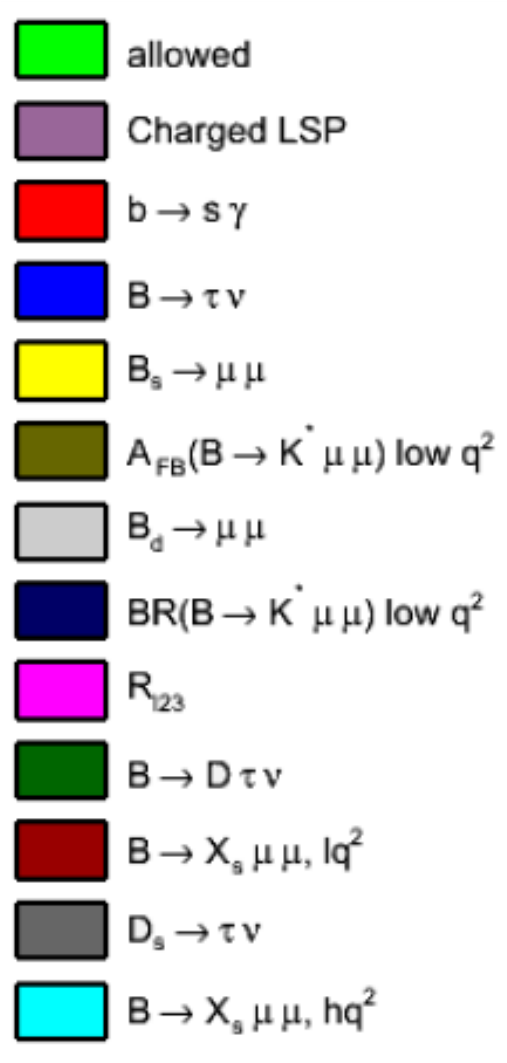}%
\caption{\label{fg:b-phys} Constraints from flavour observables in CMSSM in the plane $(m_{1/2},\,m_0)$ for $\tan\beta=30$ and $A_0 = 0$ with the 2011 results. The black line corresponds to the CMS exclusion limit with 1.1~fb$^{-1}$ of data~\cite{cms-b1} and the red line to the CMS exclusion limit with 4.4~fb$^{-1}$ of data~\cite{cms-b2}. From~\cite{b-phys}.}
\end{figure}

\subsection{SUSY direct searches at LHC}\label{sc:LHC}

SUSY searches in ATLAS~\cite{atlas-det} and CMS~\cite{cms-det} experiments typically focus on events with high transverse missing energy (\met) which can arise from (weakly interacting) LSPs, in the case of \R-parity conserving SUSY, or from neutrinos produced in LSP decays, when \R-parity is broken. Hence, the event selection criteria of inclusive channels are based on large \met, no or few leptons ($e$, $\mu$), many jets and/or $b$-jets, $\tau$-leptons and photons. In addition, kinematical variables such as the transverse mass, \mt, and the effective mass, \mef, assist in distinguishing further SUSY from SM events, whilst the \emph{effective transverse energy}~\cite{alberto-ete} can be useful to cross-check results, allowing a better and more robust identification of the SUSY mass scale, should a positive signal is found. Although the majority of the analysis simply look for an excess of events over the SM background, there is an increasing application of distribution shape fitting techniques~\cite{shape}.  

Typical SM backgrounds are top-quark production --- including single-top ---, $W$/$Z$ in association with jets, dibosons and QCD multi-jet events. These are estimated using semi- or fully data-driven techniques. Although the various analyses are motivated and optimised for a specific SUSY scenario, the interpretation of the results are extended to various SUSY models or topologies.  

Analyses exploring \R-parity conserving (RPC) SUSY models are currently divided into inclusive searches for: (a) squarks and gluinos, (b) third-generation fermions, and (c) electroweak production ($\tilde{\chi}^0$, $\tilde{\chi}^{\pm}$, $\tilde{\ell}$). Recent results from each category of CMS searches are presented in Ref.~\cite{barela}. It is stressed that, although these searches are designed to look for RPC SUSY, interpretation in terms of \R-parity violating models is also possible (cf.\ Section~\ref{sc:RPV}).

Strong SUSY production is searched in events with large jet multiplicities and large missing transverse momentum, with and without leptons. Various channels fall into this class of searches; here the 0-lepton plus three jets plus \met\ analysis from CMS is highlighted. The results are interpreted in terms of the CMSSM (Fig.~\ref{fg:cms-gl-sq}) excluding squark masses up to 1400~\gev, gluino mass up to 900~\gev, or $m_{\tilde{q}} \sim m_{\tilde{g}} \sim 1400~\gev$~\cite{cms-0lep-1}. A different approach is to present the limits in terms of a simplified topology, namely $\tilde{q}\tilde{q}\to q\tilde{\chi}_1^0 q\tilde{\chi}_1^0$, assuming that $m_{\tilde{g}} \gg m_{\tilde{q}}$~\cite{cms-0lep-2}. The results in the $(m_{\tilde{q}},\,m_{\text{LSP}})$ plane are shown in Fig.~\ref{fg:cms-simplified}.
 
\begin{figure}[ht]
\begin{minipage}[b]{0.48\textwidth}
\includegraphics[width=\textwidth]{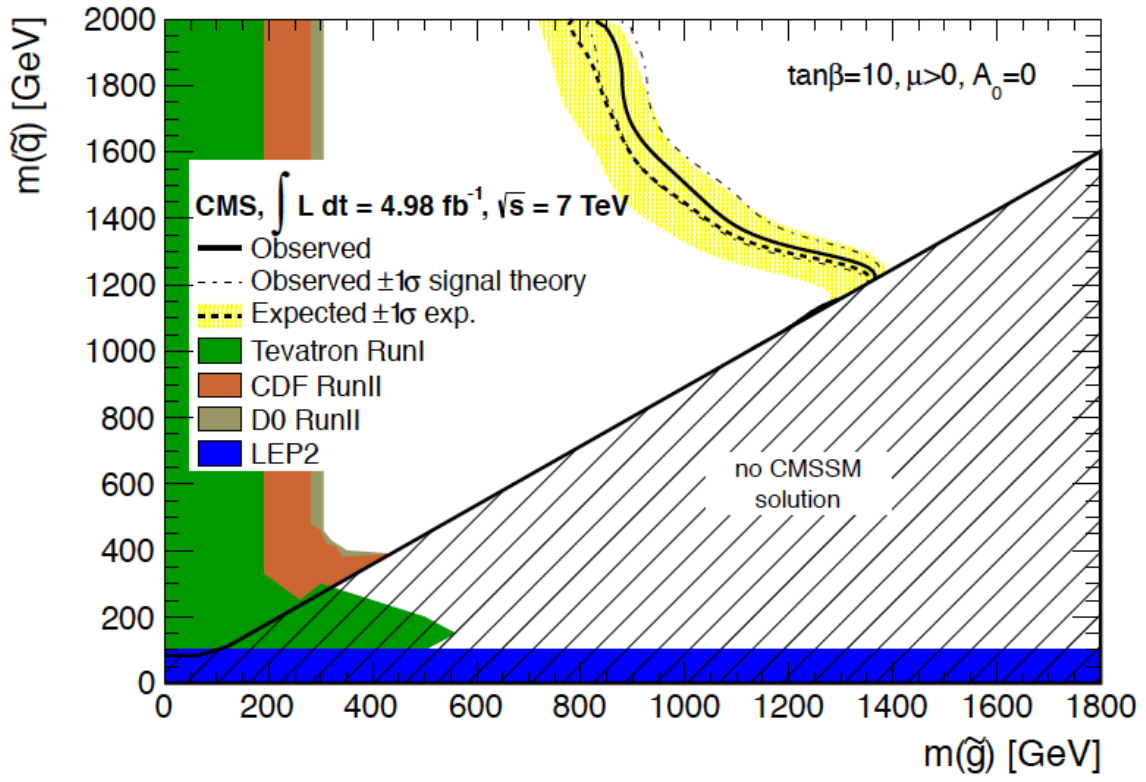}
\caption{\label{fg:cms-gl-sq} Observed and expected 95\% CL lower limits in the CMSSM $(m_{\tilde{g}},m_{\tilde{q}})$ plane, for $\tan\beta = 10$, $\mu > 0$, and $A_0 = 0$. The yellow-shaded region shows the $1\sigma$ variation in the expected limit, while the dot-dashed curves show the variation in the observed limit when the signal cross section is varied by its theoretical uncertainties. Limits from earlier searches by other experiments derived with different models or parameter choices are also shown. From~\cite{cms-0lep-1}.}
\end{minipage}\hspace{0.04\textwidth}%
\begin{minipage}[b]{0.48\textwidth}
\includegraphics[width=\textwidth]{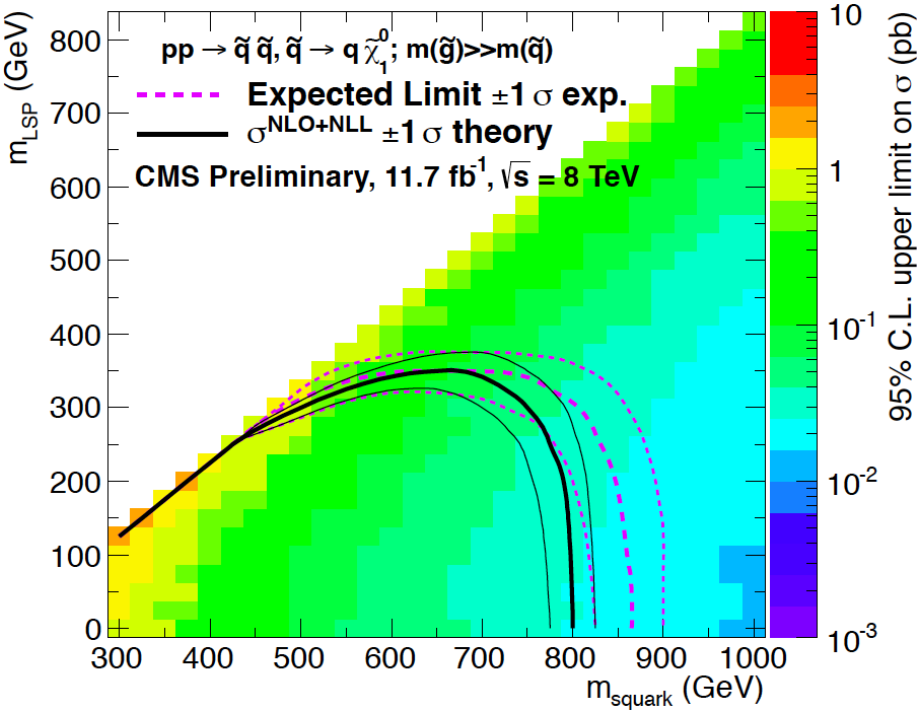}
\caption{\label{fg:cms-simplified} Upper limit on cross section at 95\%CL as a function of $m_{\tilde{q}}$ and $m_{\text{LSP}}$ for 1-step $\tilde{q}\tilde{q} \rightarrow q\tilde{\chi}_1^0q\tilde{\chi}_1^0$ simplified model. The solid thick black line indicates the observed exclusion region assuming NLO+NLL SUSY production cross section. The thin black lines represent the observed excluded region when varying the cross section by its theoretical uncertainty. The dashed purple lines indicate the median (thick line) $\pm1\sigma$ (thin lines) expected exclusion regions. From~\cite{cms-0lep-2}.}
\end{minipage} 
\end{figure}

Apart from inclusive analyses targeting phenomenological models, a wide spectrum of searches is dedicated on specific sparticle decay chains predicted in well-motivated theoretical supersymmetric models. For instance, in gauge mediated supersymmetry breaking (GMSB) models, the SUSY breaking is mediated from a hidden sector by gauge interactions. Their phenomenology is characterised by a light gravitino LSP with the NLSP mass, nature and lifetime determining the final states and therefore the analysis strategy with which to probe this model~\cite{gmsb-ifae}. A higgsino-like neutralino NLSP, for instance, preferably decays to a gravitino and a $Z$~boson or a photon. Such signatures can be detected by looking for events with a leptonically-decaying $Z$, high \met\ and jets. The ATLAS analysis~\cite{atlas-zmet} yielded the exclusion limits depicted in Fig.~\ref{fg:atlas-zmet} in the context of the general gauge mediation (GGM) model.

\begin{figure}[ht]
\begin{minipage}[b]{0.48\textwidth}
\includegraphics[width=\textwidth]{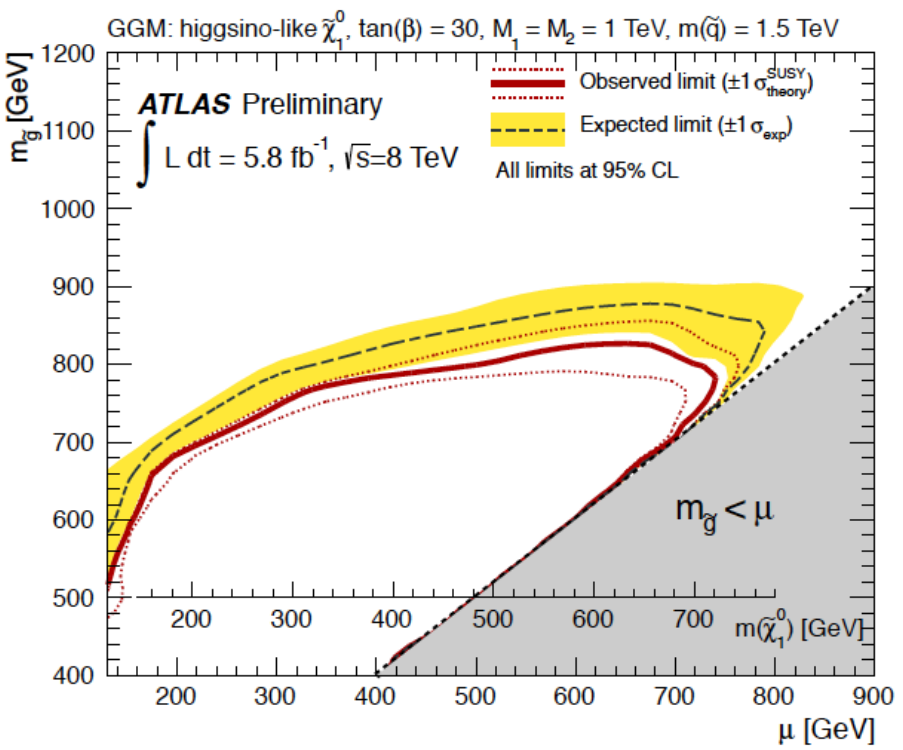}
\caption{\label{fg:atlas-zmet}Expected and observed 95\% CL exclusion limits for SR1 on the $m_{\tilde{g}}$ and $\mu$ parameters for GGM models with $\tan\beta=30$, $M_1=M_2=1$~TeV, $c\tau_{\text{NLSP}} < 0.1$~mm, and $m_{\tilde{q}}=1.5$~TeV. The light grey area indicates the region where the NLSP is the gluino, which is not considered in this analysis. From~\cite{atlas-zmet}.}
\end{minipage}\hspace{0.04\textwidth}%
\begin{minipage}[b]{0.48\textwidth}
\includegraphics[width=\textwidth]{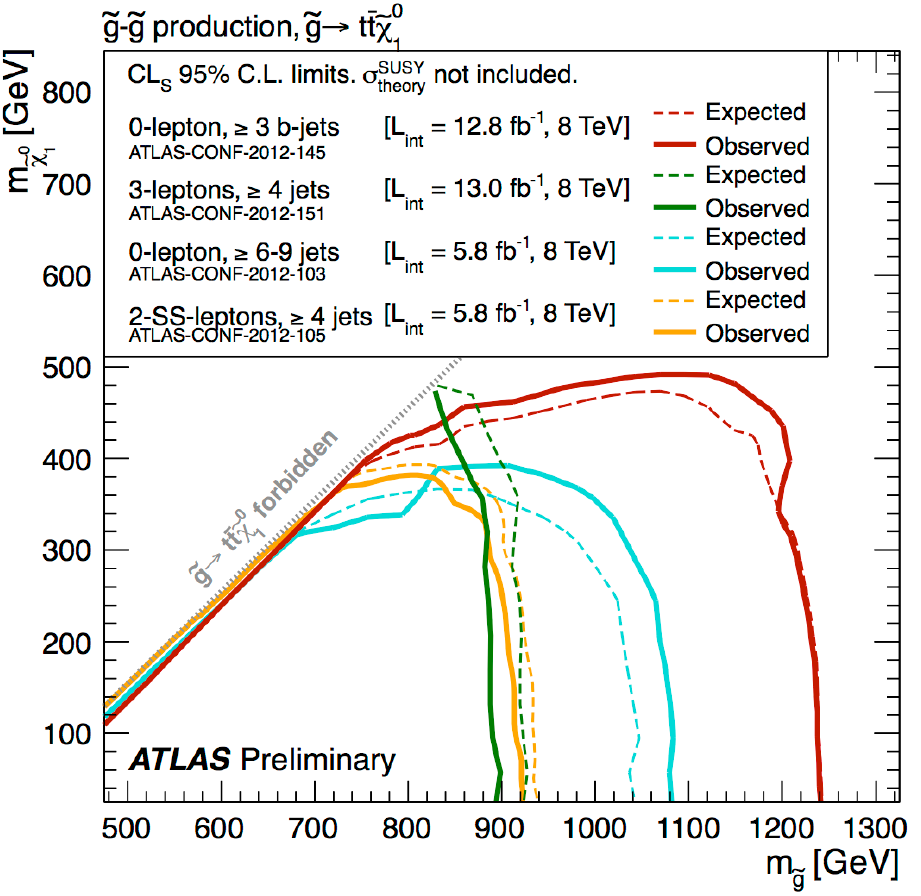}
\caption{\label{fg:atlas-stop}Exclusion limits at 95\% CL for 8~TeV analyses in the $(m_{\tilde{g}},m_{\tilde{\chi}_1^0})$ plane for the Gtt simplified model where a pair of gluinos decays via off-shell stop to four top quarks and two neutralinos (LSP). From~\cite{atlas-stop}.}
\end{minipage} 
\end{figure}

The mixing of left- and right-handed gauge states which provides the mass eigenstates of the scalar quarks and leptons can lead to relatively light 3$^\mathrm{rd}$ generation particles. Stop ($\tilde{t}_1$) and sbottom ($\tilde{b}_1$) with a sub-TeV mass are favoured by the naturalness argument, while the stau ($\tilde{\tau}_1$) is the lightest slepton in many models. Therefore these could be abundantly produced either directly or through  gluino production and decay. Such events are characterised by several energetic jets (some of them $b$-jets), possibly accompanied by light leptons, as well as high \met. 

Both main LHC experiments, ATLAS and CMS, have embarked in analyses dedicated to unveil third-generation sparticle production covering a wide spectrum of signatures. The results extracted by ATLAS with up to 13~\ifb\ of data at $\sqrt{s}=8~\tev$, as far as the gluino-mediated production is concerned, are summarised in Fig.~\ref{fg:atlas-stop}. Gluino masses of up to 1240~\gev\ and lightest neutralino masses up to 500~\gev\ in the Gtt simplified model have been excluded. More updated results on the third-generation are available at Refs.~\cite{cms-results,atlas-stop}, for CMS and ATLAS, respectively.  

The electroweak SUSY production at LHC proceeds through weak gaugino (charginos and neutralino) or slepton pair production and is typified by low cross section compared to the strong processes. These events are characterised by multiple charged leptons and moderately high \met. Their kinematics is quite different from the long cascade decays typically found in SUSY processes through strong interactions. Therefore, kinematic variables other than the effective mass may be deployed such as the \emph{visible transverse energy/momentum}, to improve the power of the analysis and also give way to the determination of a combination of sparticle masses~\cite{alberto-ew}. 

In certain SUSY breaking scenarios, characteristic signatures are expected involving heavy sparticles with a long lifetime. A non-exhaustive list of examples of such (meta-)stable particles includes long-lived sleptons in GMSB models and R-hadrons with long-lived gluinos or squarks in split SUSY. These particles traverse the entire detector and, due to their high mass, they move slowly ($\beta<1$), yielding high specific ionisation energy loss, ${\rm d}E/{\rm d}x$, and a long time of flight (TOF). The result of a search for heavy long-lived charged particles produced in $pp$ collisions at $\sqrt{8~\tev}$ with 5~\ifb\ of CMS data~\cite{cms-ll-hsp} is shown in Fig.~\ref{fg:cms-hsp}. The inner tracking detectors were used to define a sample of events containing high-\pt\ tracks and high ${\rm d}E/{\rm d}x$. A second sample of events, which have high-momentum tracks satisfying muon identification requirements in addition to meeting high-ionisation and long time-of-flight requirements, was analysed independently. In both cases, the results are consistent with the expected background estimated from data, thus establishing cross-section limits as a function of mass within the context of models with long-lived gluinos, scalar top quarks and scalar taus. Lower limits at 95\% confidence level on the mass of gluinos (stops) are found to be 1098 (737)~\gev. A limit of 928 (626)~\gev\ is set for a gluino (stop) that hadronizes into a neutral bound state before reaching the muon detectors. The lower mass limit for a pair produced $\tilde{\tau}$ is found to be 223~\gev. It is worthy to mention that the MoEDAL~\cite{moedal} experiment at LHC is specifically designed to explore such high-ionisation signatures.

\begin{figure}[ht]
\begin{minipage}[b]{0.42\textwidth}
\includegraphics[width=\textwidth]{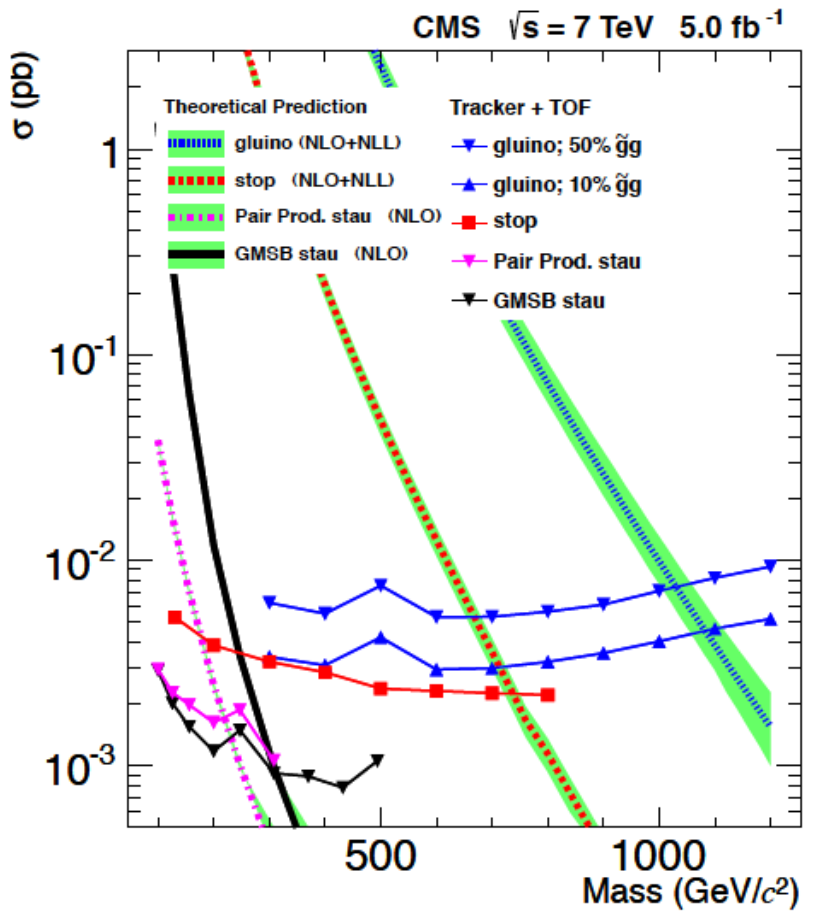}
\caption{\label{fg:cms-hsp} Predicted theoretical cross sections with associated uncertainties and observed 95\% CL upper limits on the cross section for the different signal models considered: production $\tilde{t}_1$, $\tilde{g}$, and $\tilde{\tau}_1$; different fractions of R-gluonball states, using the tracker and TOF information. From~\cite{cms-ll-hsp}.}
\end{minipage}\hspace{0.04\textwidth}%
\begin{minipage}[b]{0.54\textwidth}
\includegraphics[width=\textwidth]{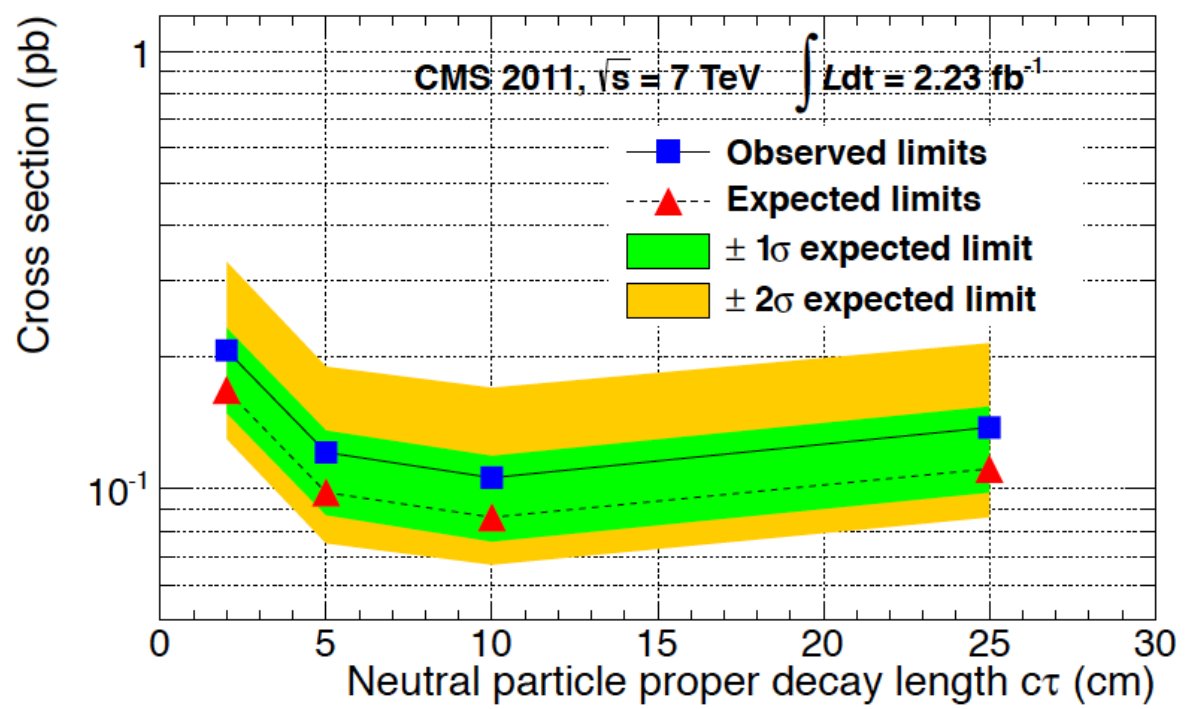}
\caption{\label{fg:cms-photons} 95\% CL upper limits on the pair production cross section for neutral particles, each of which decays into a photon and invisible particles, as a function of the neutral particle proper decay length. The observed values as a function of mass are shown by the solid line. The dashed line indicates the expected median of results for the background-only hypothesis, while the green (dark) and yellow (light) bands indicate the ranges that are expected to contain 68\% and 95\% of all observed excursions from the median, respectively. From~\cite{cms-ll-photons}.}
\end{minipage} 
\end{figure}

GMSB scenarios, on the other hand, predict signatures involving high-\pt\ photons which may or may not point back to the primary collision vertex. These final states arise in the decay of a long-lived $\tilde{\chi}_1^0$ decaying to a photon and a gravitino. CMS carried out a search using events containing photons, \met, and jets with 2.23~\ifb\ of data at $\sqrt{s}=7~\tev$~\cite{cms-ll-photons}. The impact parameter of the photon relative to the beam-beam collision point can be reconstructed using converted photons. The method is sensitive to lifetimes of the order of 0.1 to 1~ns. Cross-section limits on pair production for such particles, each of which decays into a photon and invisible particles, are presented in Fig.~\ref{fg:cms-photons}. The observed 95\% CL limits vary between 0.11 and 0.21~pb, depending on the neutral particle lifetime. 

The spectrum of SUSY searches at the LHC is huge both in terms of the variety of tested models as well as the  explored signatures and it was merely highlighted here. A summary of the SUSY-related analysis results from CMS~\cite{cms-results} is shown in Fig.~\ref{fg:cms-summary}. A similar synopsis of results from the ATLAS experiment can be found in Ref.~\cite{atlas-stop}. In conclusion, LHC probes supersymmetry in scales between 100~\gev\ and 1~\tev, without having seen yet some evidence of its existence. Experiments in upcoming $e^+e^-$ colliders, such as the ILC~\cite{ilc} and CLIC~\cite{clic}, are expected to discover and constrain SUSY in the future. 

\begin{figure}[ht]
\centering
\includegraphics[width=0.5\textwidth]{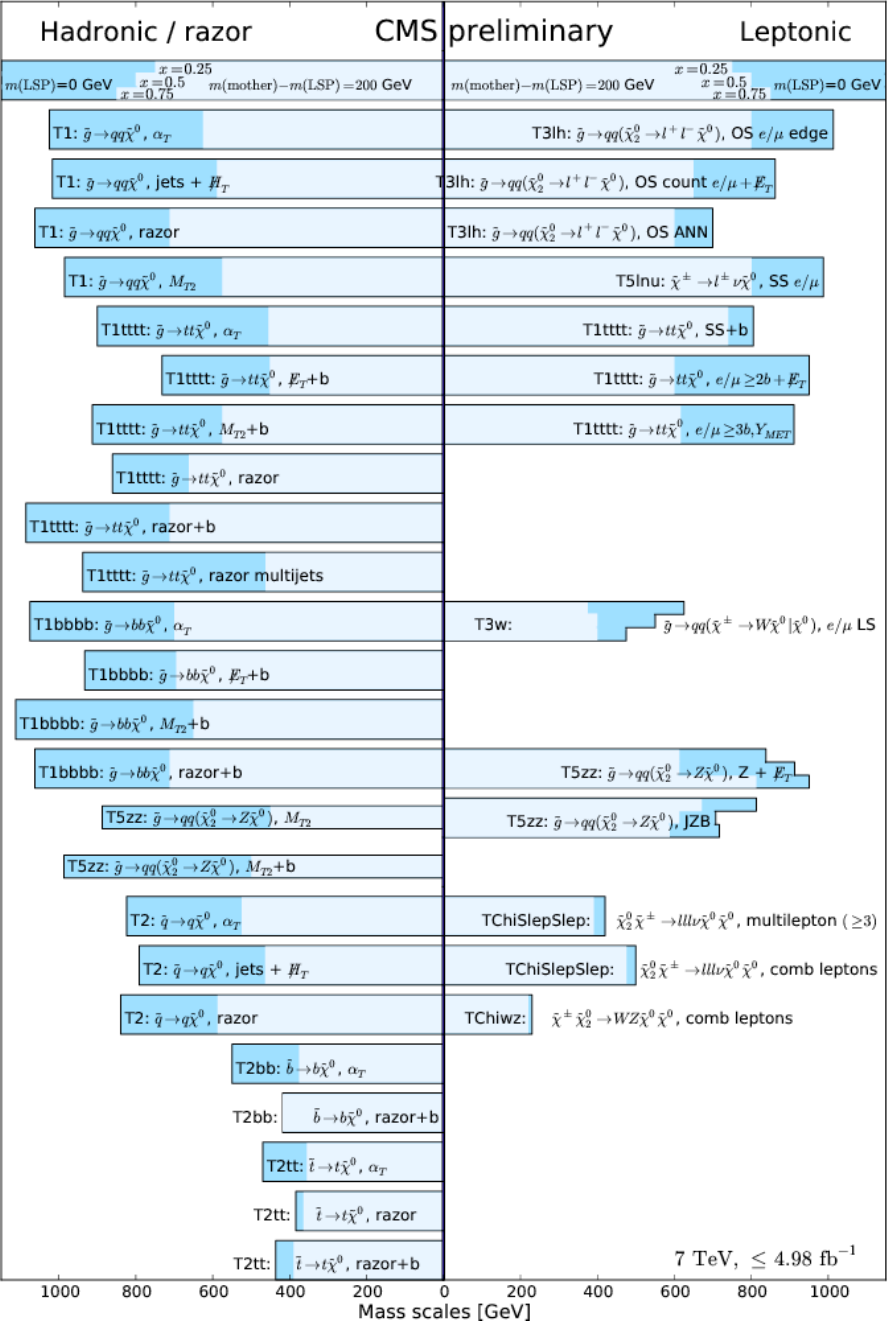}%
\caption{\label{fg:cms-summary} Exclusion limits for the masses of the mother particles, for $m_{\X} = 0~\gev$ (dark blue) and $m_{\text{mother}}-m_{\X} = 200~\gev$ (light blue), for each analysis, for the hadronic (left) and leptonic (right) results. In this plot, the lowest mass range is $m_{\text{mother}}=0$, but results are available starting from a certain mass depending on the analyses and topologies. From~\cite{cms-results}.}
\end{figure}

\subsection{Implications from the Higgs boson}\label{sc:higgs}

The recent discovery by ATLAS~\cite{atlas-higgs,goncalo} and CMS~\cite{cms-higgs} of a new boson with properties consistent with those of a SM-like Higgs boson shed new light in the supersymmetric landscape~\cite{djouadi}. The newly observed boson having a mass of $~125-126~\gev$ is rather heavy for being the lightest supersymmetric Higss, $h$, however it still lies in the allowed range. If $h$ is SM-like, i.e.\ in the decoupling limit, the heaviest neutral SUSY Higgs bosons, $H$ and $A$ are pushed to the $1~\tev$ scale. In this case, high values of $\tan\beta$ are favoured, together with high stop mixing and heavy squarks. 

Some (minimal) models are disfavoured, while other remain viable even with other constraints present, e.g.\ the NMSSM~\cite{nmssm}. The singlet that is introduced in this model to evade the $\mu$ problem, leads to a richer Higgs sector than the MSSM that can accommodate a relatively heavy Higgs mass~\cite{nmssm-higgs}. This is demonstrated in Fig.~\ref{fg:higgs}, where the blue points denote the models with Higgs boson masses compatible with the measured Higgs mass of $125~\gev$ within $\pm 3~\gev$. Moreover the black points also reproduce the observed $\gamma\gamma$ event rate.  

\begin{figure}[ht]
\includegraphics[width=0.47\textwidth]{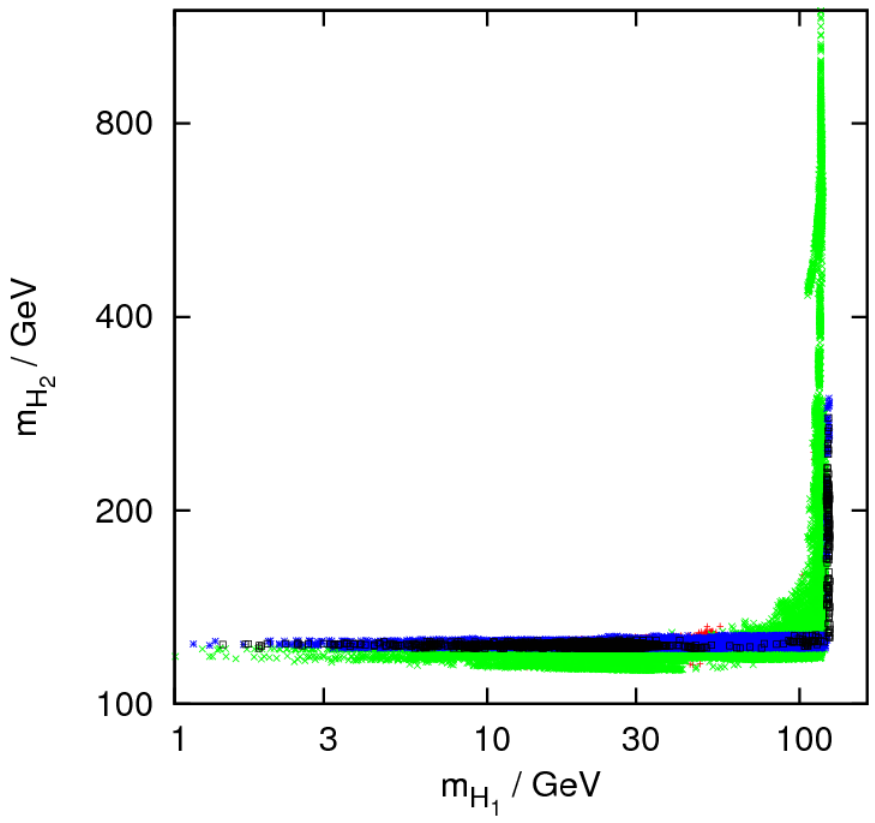}
\hspace{2pc}%
\includegraphics[width=0.47\textwidth]{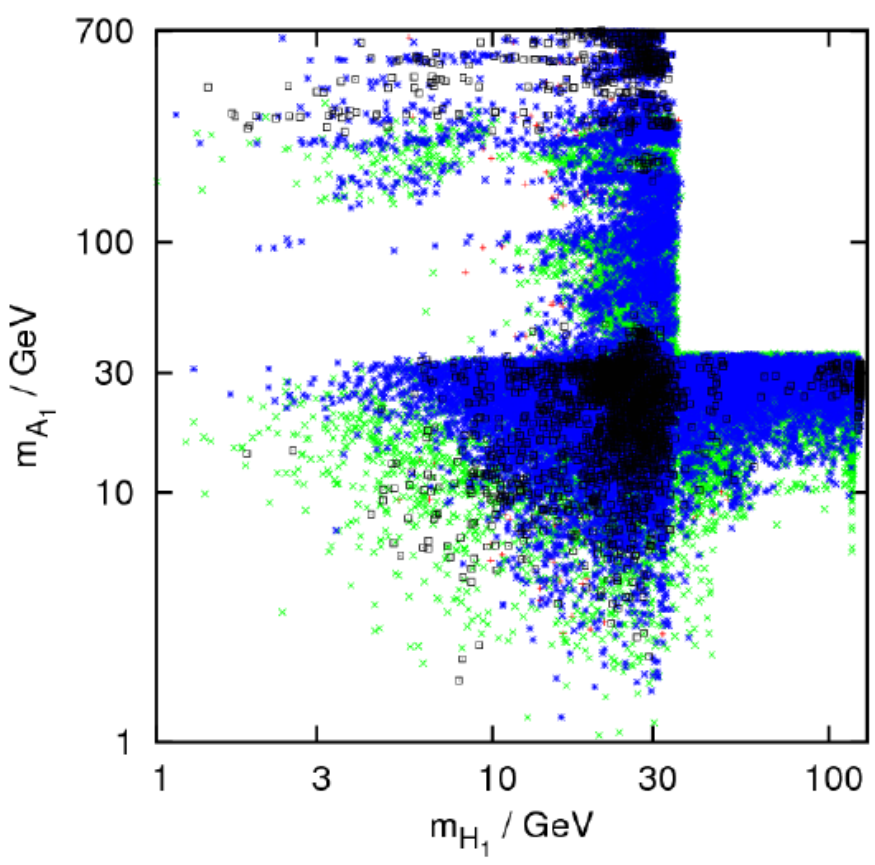}
\caption{\label{fg:higgs} Masses of the Higgs (pseudo-)scalars $H_1,H_2$ (left) and $H_1,A_1$ (right). Red points are ruled out either by {\tt HiggsBounds} constraints or the ATLAS 1~\ifb\ jets and \met\ SUSY search. Green points have no Higgs with a mass in the $122-128~\gev$ interval, blue points have a Higgs ($H_1$ and/or $H_2$) within this mass range, and black points have a Higgs such that $R_{gg\gamma\gamma} > 0.4$. From~\cite{nmssm-higgs}.}
\end{figure}

\subsection{\R-parity violation}\label{sc:RPV}

Throughout the previous discussion, we assumed that \R-parity --- a multiplicative quantum number defined as $R=(-1)^{3(B-L)+2s}$, where $B$, $L$ and $s$ are the baryon number, the lepton number and the spin, respectively --- was conserved. This supposition is hinted but not required by proton stability; rapid proton decay is avoided if one of $B$ or $L$ is conserved. Allowing \R-parity violating (RPV) terms~\cite{rpv} in the superpotential has decisive implications for SUSY phenomenology in colliders. First of all, single sparticle production is possible. The LSP may be charged and/or carry colour and, most importantly, decays leaving open the possibility for new discovery signatures and for the LSP-mass reconstruction through the invariant mass of its decay products. Furthermore, the \met\ may or may not be high depending on the LSP-decay products and the underlying RPV model.   

Among the various ways to break \R-parity, the violation through bilinear terms is especially interesting for its connection with neutrino physics. In this model the spontaneous breaking of \R-parity gives vacuum expectation values (vevs) to sneutrinos providing a \emph{vev-seesaw} mechanism that leads to neutrino masses~\cite{brpv}. Below the scale of these vevs, \R-parity breaking is explicit through bilinear $L$-violating terms and their corresponding soft breaking terms. The same parameters that induce neutrino masses and mixings are also responsible for the decay of the LSP which induces a relation between some decay modes of the LSP and neutrino mixing angles. The latest exclusion limits set on this model, when the bilinear terms are embedded in mSUGRA, were obtained with 5~\ifb\ of ATLAS data at $\sqrt{s}=7~\tev$ by searching for events with high jet multiplicity, \met\ and one lepton~\cite{atlas-brpv2} and are shown in Fig.~\ref{fg:atlas-brpv}.

\begin{figure}[ht]
\begin{minipage}[b]{0.48\textwidth}
\includegraphics[width=0.98\textwidth]{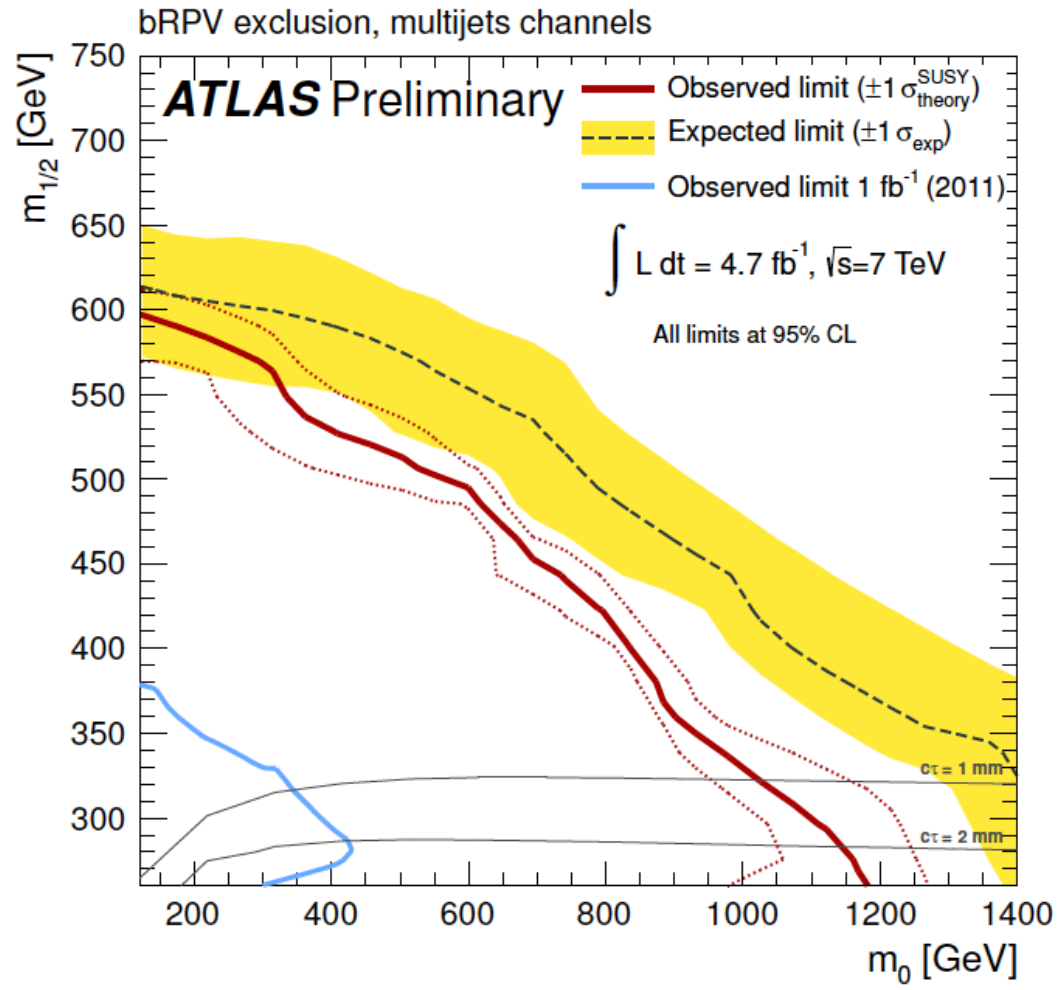}
\caption{\label{fg:atlas-brpv} Expected and observed 95\% CL exclusion limits in the bilinear \R-parity violating model obtained by combining the $e$ and $\mu$ channels. The band around the median expected limit shows the $\pm1\sigma$ variations on the median expected limit, including all uncertainties except theoretical uncertainties on the signal. The dotted lines around the observed limit indicate the sensitivity to $\pm1\sigma$ variations on these theoretical uncertainties. The thin solid black contours show the LSP lifetime. The result from the previous ATLAS search~\cite{atlas-brpv1} is also shown. From~\cite{atlas-brpv2}.}
\end{minipage}\hspace{0.04\textwidth}%
\begin{minipage}[b]{0.48\textwidth}
\includegraphics[width=0.98\textwidth]{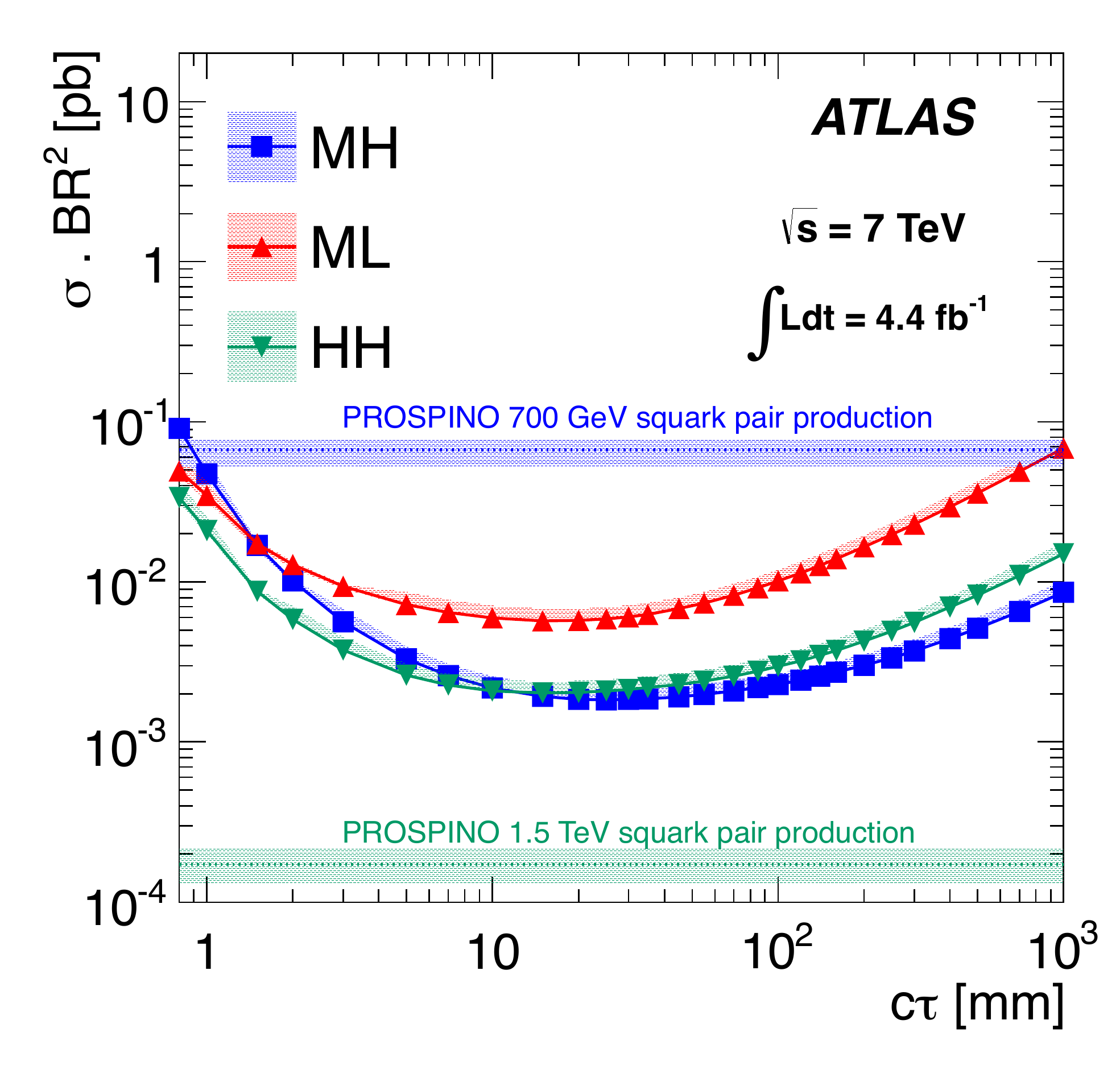}
\caption{\label{fg:atlas-dv} Upper limits at 95\% CL on $\sigma\cdot BR^2$ vs.\ the neutralino lifetime for different combinations of squark and neutralino masses, based on the observation of zero events satisfying all criteria in a 4.4~\ifb\ data sample. The shaded areas around these curves represent the $\pm1\sigma$ uncertainty bands on the expected limits. The horizontal lines show the cross sections calculated from PROSPINO for squark masses of 700~\gev\ and 1500~\gev. The shaded regions around these lines represent the uncertainties on the cross sections. From~\cite{atlas-dv}.}
\end{minipage} 
\end{figure}

Moreover, the potentially long LSP lifetime may give rise to displaced vertices (DV) in the detector. For instance, an RPV SUSY scenario, where the non-zero RPV coupling $\lambda'_{2ij}$ allows the decay $\tilde{\chi}_1^0\to\mu q\bar{q}' $, would give rise to a multi-track DV that contains a high-\pt\ muon at a distance between millimeters and tens of centimeters from the $pp$ interaction point. The results of such a search~\cite{atlas-dv} performed by ATLAS are shown in Fig.~\ref{fg:atlas-dv}. Fewer than 0.03 background events are expected in the data sample of 33~\ipb, and no events are observed. Based on this null observation, upper limits are set on the supersymmetry production cross-section $\sigma\times B$ of the simulated signal decay chain for different combinations of squark and neutralino masses and for different values of the neutralino lifetime, $c\tau$.

The occurrence of DVs is inherent in the ``$\mu$ from $\nu$'' supersymmetric standard model ($\mu\nu$SSM)~\cite{munussm}, which solves the naturalness problem of the MSSM, the so-called $\mu$-problem, and explains  the origin of neutrino masses by simply using right-handed neutrino superfields. The model is characterised by three singlets (as opposed to one in NMSSM) and bilinear RPV couplings, giving rise to a rich phenomenology with many Higgs bosons and neutralinos. It has been demonstrated~\cite{munussm-dv} that a 125-\gev\ Higgs boson while decaying to a pair of unstable long-lived neutralinos, can lead to a distinct signal with non-prompt multileptons. This signal provides an unmistakable signature of the model, pronounced with light neutralinos. Evidence of this signal is well envisaged with sophisticated displaced vertex analysis, as shown in Fig.~\ref{fg:munussm-dv}.

\begin{figure}[ht]
\begin{minipage}[b]{0.48\textwidth}
\includegraphics[width=\textwidth]{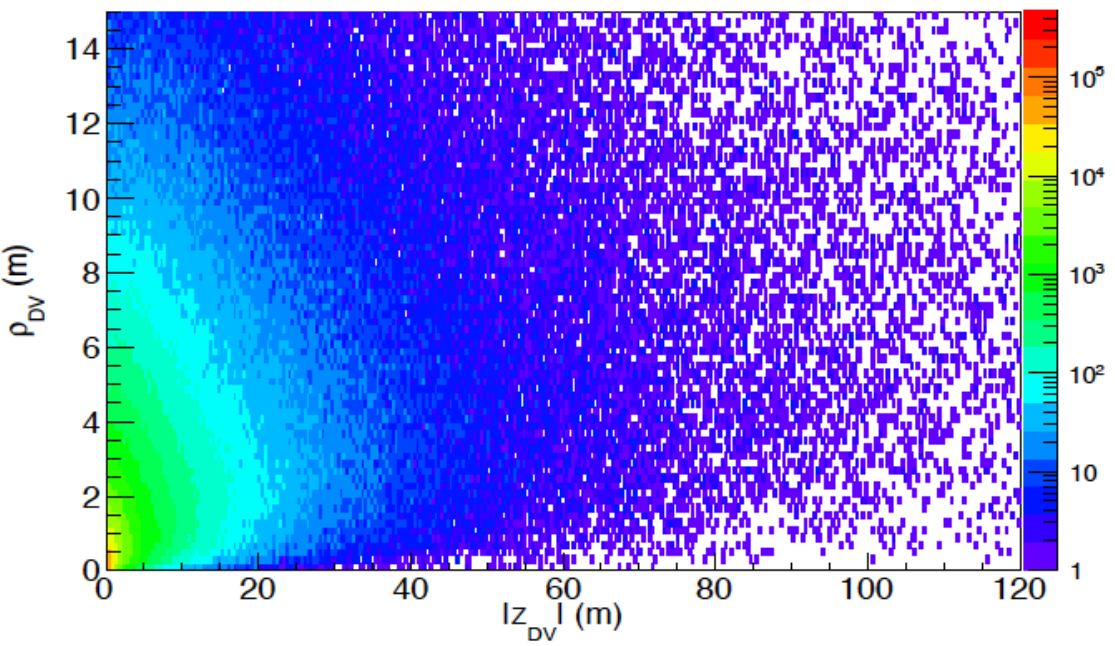}
\caption{\label{fg:munussm-dv} Cylindrical coordinates $\rho_{\text{DV}}$ versus $|z_{\text{DV}}|$ for the $\tilde{\chi}_4^0 \to \tau\tau\nu$ decay, when a $\mu\nu$SSM Higgs is produced at $pp$ collisions at $\sqrt{s} = 8~\tev$ and an integrated luminosity of $20~\ifb$. From~\cite{munussm-dv}.}
\end{minipage}\hspace{0.04\textwidth}%
\begin{minipage}[b]{0.48\textwidth}
\includegraphics[width=\textwidth]{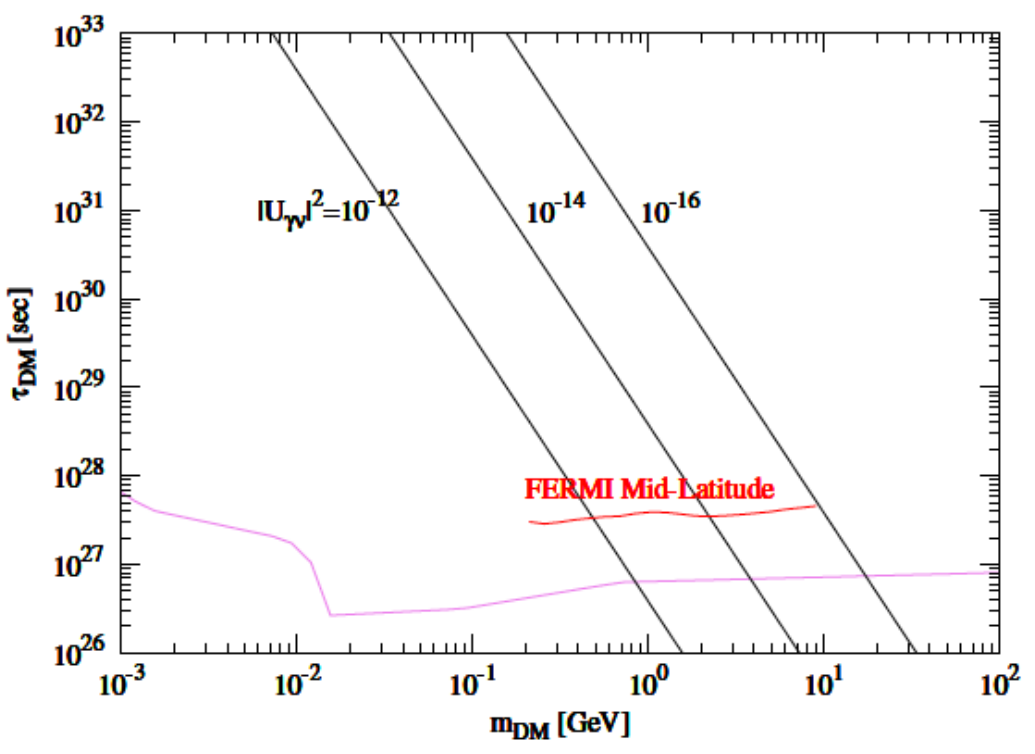}
\caption{\label{fg:munussm-dm} Constraints on $\tilde{G}$ DM lifetime vs.\ mass in $\mu\nu$SSM. The region below the magenta (red) line is excluded by $\gamma$-ray observations (Fermi). Black lines denote predictions for several values of $|U_{\tilde{\gamma}\nu}|^2$. From~\cite{munussm-dm}.}
\end{minipage} 
\end{figure}

As a last remark, we address the issue of (not necessarily cold) dark matter in RPV SUSY models. These seemingly incompatible concepts \emph{can} be reconciled in models with a gravitino~\cite{rpv-grav} or an axino~\cite{rpv-axino} LSP with a lifetime exceeding the age of the Universe. In both cases, RPV is induced by bilinear terms in the superpotential that can also explain current data on neutrino masses and mixings without invoking any GUT-scale physics. Decays of the next-to-lightest superparticle occur rapidly via RPV interaction, and thus they do not upset the Big-Bang nucleosynthesis, unlike the \R-parity conserving case. Such gravitino DM is proposed in the context of $\mu\nu$SSM with profound prospects for detecting $\gamma$ rays from their decay~\cite{munussm-dm}. The constraints on $\tilde{G}$ lifetime versus mass set by Fermi and $\gamma$-ray observations are summarised in Fig.~\ref{fg:munussm-dm}. Values of the gravitino mass larger than 10~\gev\ are disfavoured, as well as lifetimes smaller than about $3$ to $5\times10^{27}~{\rm s}$.

\section{Summary}\label{sc:summary}

The experimental and observational landscape of SUSY searches is multifaceted. In the dark matter front, the cosmological evidence helps orientate the SUSY model building, while the exploration through direct and indirect detection continues with intriguing results. Concerning the high-energy colliders, the LHC is in the forefront: powerful direct searches probe various manifestations of SUSY, whilst indirect information is collected. No signal has been observed so far, thus strong limits have been set up to a SUSY mass scale of $\sim 1~\tev$. The search strategies keep evolving to explore the most theoretically-attractive models and new ideas. In both Particle and Astroparticle fronts, the quest for Supersymmetry continues; many possibilities are still open.

\ack
The author is grateful to the DISCRETE2012 organisers for the kind invitation and support. Thanks to them, a warm, friendly and intellectually stimulating atmosphere was enjoyed by the speakers and participants throughout the Symposium. This work was supported in part by the Spanish Ministry of Economy and Competitiveness (MINECO) under the project FPA2009-13234-C04-01 and by the Spanish Agency of International Cooperation for Development under the PCI project A1/035250/11.


\section*{References}


\end{document}